\documentclass[12pt]{iopart}

 \usepackage{graphics}
 \usepackage{epsfig}
\begin{document}

\title[Thermodynamic quantum critical behavior of the anisotropic Kondo necklace model]{Thermodynamic  quantum critical behavior of the anisotropic Kondo necklace model}

\author{D. Reyes$^1$, M. A. Continentino$^2$, Han-Ting Wang$^3$}

\address{$^1$Centro Brasileiro de Pesquisas F\'{\i}sicas - Rua Dr.
Xavier Sigaud, 150-Urca,
 \\ 22290-180,RJ-Brazil}
\address{$^2$Instituto de F\'{\i}sica, Universidade Federal Fluminense, \\
Campus da Praia Vermelha, \\
Niter\'oi, RJ, 24.210-340, Brazil}
\address{$^3$Beijing National Laboratory of Condensed Matter Physics and Institute of Physics,\\ Chinese Academy of Sciences,\\ Beijing 100080,
People's Republic of China}
\ead{\mailto{daniel@cbpf.br}}

\begin{abstract}

The Ising-like anisotropy parameter $\delta$ in the Kondo necklace model is analyzed using the 	
bond-operator method at zero and finite temperatures for arbitrary $d$ dimensions.
A decoupling scheme on the double time Green's functions is used to find the dispersion
relation for the excitations of the system. At zero temperature and in the paramagnetic side of the phase diagram, we determine the spin gap exponent
$\nu z\approx0.5$ in three dimensions and anisotropy between $0\leq\delta\leq1$, a result consistent with the dynamic exponent $z=1$ for the Gaussian character of the bond-operator treatment.
At low but finite temperatures, in the antiferromagnetic phase, the line of Neel transitions is calculated for $\delta\ll1$ and $\delta\approx1$. For $d>2$ it is only re-normalized by the anisotropy parameter and  varies with the distance to the quantum critical point QCP $|g|$ as, $T_N \propto |g|^{\psi}$ where the shift exponent $\psi=1/(d-1)$. Nevertheless, in two dimensions, long range magnetic order occurs only at $T=0$ for any $\delta$.
In the paramagnetic phase, we find a power law temperature dependence on the specific heat at the \textit{quantum liquid trajectory} $J/t=(J/t)_{c}$, $T\rightarrow0$. It behaves as $C_{V}\propto T^{d}$ for $\delta\leq 1$ and $\delta\approx1$, in concordance with the scaling theory for $z=1$.

\end{abstract}
\maketitle

\section{Introduction}
Quantum phase transitions (QPT) from an antiferromagnetic
$AF$ ordered state to a nonmagnetic Fermi liquid (NFL) in heavy
fermion (HF) systems have received considerable attention from both theoretical\cite{D2} and experimental points of view\cite{Julio}. In contrast to classical phase transitions (CPT), driven by
temperature, QPT can be driven by tuning an independent-temperature control parameter
(magnetic field, external pressure, or doping). The physics of HF is mainly due to the competition
of two main effects: the Ruderman-Kittel-Kasuya-Yosida
(RKKY) interaction between the magnetic ions which favors
long-range magnetic order and the Kondo effect which tends to screen the local moments and produce a nonmagnetic ground
state. These effects are contained in the Kondo lattice model
(KLM) Hamiltonian in which, only spin degrees of freedom are considered.
Here we investigated a simplified version, the so-called Kondo necklace model\cite{Doniach} (KNM) which for all
purposes can be considered yield results similar to the original model. While the ground state properties of this model has been investigated rather extensively, by a variety of methods\cite{Matsu,Kotov,Scalettar,Jullien,Santini,Moukouri,Otsuka,Zhang,Langari,Strong,Kiselev}, thermodynamic and finite temperature critical properties, close to a magnetic instability, remain an open issue. That was warned for us, and it was our first motivation for studying the quantum critical properties of this model, as a function of the distance to the quantum critical point $|g|$ at zero and low temperatures\cite{D2,D1}. We extend now this treatment to finite inter-site anisotropy $\delta$ such that, $0\leq\delta\leq1$ since
the $\delta=1$ case is appropriate to describe compounds where the ordered magnetic phase has a strong Ising component. However, the main reason for considering anisotropy $\delta$ in the KNM is to try to describe its effects in the neighborhood of a magnetic quantum critical point (QCP) in HF systems, rather than a symmetry problem\cite{Langari}. This is a goal in HF systems, and already several theories were formulated to explain their unusual properties\cite{Mucio1,Moriya,Hertz}. Besides, in an early work we were succeeded in finding that the Neel line
exists since turning on a geometric anisotropy\cite{D3}, stressing that anisotropy is an inherent ingredient in real HF systems.
Henceforth, the model will be called anisotropic Kondo necklace model (AKNM). This model was already investigated using the real space
renormalization group machinary\cite{Saguia} but just in one dimension and zero temperature.
We use the bond-operator approach introduced by Sachdev and Bhatt\cite{sachdev} which was employed previously to both,
KLM\cite{jure} and KNM\cite{Zhang} models but always at $(T,\delta)=(0,0)$.
We find that this method yields a \emph{shift exponent} that characterizes the shape of the critical
line in the neighborhood of the QCP, as well as, the power
law temperature dependence on the specific heat along the so-called \textit{quantum critical trajectory} $J/t=(J/t)_c$, $T\rightarrow0$.
We consider the following AKNM:
\begin{equation}\label{AKNM}
H=t\sum_{<i,j>}(\tau^{x}_{i}\tau^{x}_{j}+(1-\delta)\tau^{y}_{i}\tau^{y}_{j})+J\sum_{i}\mathbf{S}_{i}.\mathbf{\tau}_{i},
\end{equation}
where $\tau_{i}$ and $\mathbf{S}_{i}$ are independent sets of
spin-1/2 Pauli operators, representing the conduction electron spin
and localized spin operators, respectively. The sum $\langle
i,j\rangle$ denotes summation over the nearest-neighbor sites. The
first term mimics electron propagation which strength $t$ and the second term is the magnetic interaction between
conduction electrons and localized spins $\mathbf{S}_{i}$ via the Kondo exchange coupling $J$ $(J>0)$. The Ising-like anisotropy parameter $\delta$ varies from the full anisotropic case $\delta=1$ to the
well established case $\delta=0$.
Considering the bond-operator representation for two
spins $S=1/2$, $\tau_{i}(S_{i})^{\alpha}=\mp\frac{1}{2}(s_i^{\dagger }t_{i,\alpha
}+t_{i,\alpha }^{\dagger }s_i\pm i\epsilon _{\alpha \beta \gamma
}t_{i,\beta }^{\dagger }t_{i,\gamma })$ $(\alpha=x,y,z)$\cite{sachdev},
the Hamiltonian above, at half-filling, i.e., with one conduction
electron per site, can be simplified and the resulting effective Hamiltonian $H_{mf}$ with only quadratic
operators is sufficient to describe exactly the quantum phase
transition from the disordered Kondo spin liquid to the AF phase, as discussed below.
Then, we have a mean-field Hamiltonian:

\begin{eqnarray}\label{mfAKNM}
H_{mf}&=&N\left( -\frac 34J\overline{s}^2+\mu
\overline{s}^2-\mu \right)+\omega_{0} \sum_{{\bf k}}t_{{\bf
k},z}^{\dagger
}t_{{\bf k},z}^{\phantom\dagger}\nonumber\\
 &+&\sum_{\bf k}\left[ \Lambda _{{\bf k}}t_{{\bf k},x }^{\dagger
}t_{{\bf k},x}^{\phantom\dagger}+\Delta _{{\bf k}}\left( t_{{\bf k}
,x }^{\dagger }t_{-{\bf k},x }^{\dagger }+t_{{\bf k},x }t_{-{\bf
k},x
}\right) \right]\nonumber\\
&+&\sum_{\bf k}\left[ \Lambda _{{\bf k}}'t_{{\bf k},y }^{\dagger
}t_{{\bf k},y}+\Delta _{{\bf k}}'\left( t_{{\bf k} ,y }^{\dagger
}t_{-{\bf k},y }^{\dagger }+t_{{\bf k},y }t_{-{\bf k},y
}\right) \right],
\end{eqnarray}
where $\Lambda _{{\bf k}}=\omega_{0}+2\Delta_{{\bf k}}$, $\Lambda
_{{\bf k}}'=\omega_{0}+2\Delta_{{\bf k}}'$, $\Delta _{{\bf k}}=\frac
14t\overline{s}^2\lambda ( {\bf k)}$, $\Delta _{{\bf k}}'=\frac
14t\overline{s}^2\lambda ({\bf k)}(1-\delta)$ and $\lambda ({\bf
k)=}\sum_{s=1}^d\cos k_s$. $\overline{s}$ is the singlet order
parameter consistent with the strong coupling limit $J/t\rightarrow \infty$, where the model becomes trivial, since each $\bf{S}$ spin captures a conduction electron spin to form a singlet, and where the ground state corresponds to a direct product of those singlets. The chemical potential $\mu$ was introduced to impose the constraint condition of single occupancy,  $N$ is the number of lattice sites and $Z$ is the total number of the nearest neighbors on the hyper-cubic
lattice. The wavevectors $k$ are taken in the first Brillouin zone
and the lattice spacing was assumed to be unity. This mean-field
Hamiltonian can be solved using the Green's functions to obtain the
thermal averages of the singlet and triplet correlation functions.
These are given by,

\begin{eqnarray}\label{propag}
\ll
t_{{\bf k},x}^{\phantom\dagger};t_{{\bf k},x}^{\dag}\gg&=&\frac{(\omega^{2}-\omega_{k}'^{2})(\omega+\Lambda_{k})}{2\pi\xi},\hspace{2mm}\nonumber\\
\ll
t_{{\bf k},y}^{\phantom\dagger};t_{{\bf k},y}^{\dag}\gg&=&\frac{(\omega^{2}-\omega_{k}^{2})(\omega+\Lambda_{k}')}{2\pi\xi},\hspace{2mm}\nonumber\\
\ll t_{{\bf k},z}^{\phantom\dagger};t_{{\bf
k},z}^{\dag}\gg&=&\frac{1}{2\pi(\omega-\omega_{0})},\
\end{eqnarray}
where $\xi=(\omega^{2}-\omega_{k}^{2})(\omega^{2}-\omega_{k}'^{2})$. The poles of the Green's functions determine the excitation energies
of the system as $\omega_{0}=\left(\frac{J}{4}+\mu\right)$, which is
the dispersionless spectrum of the longitudinal spin triplet states,
$\omega_{k}=\pm \sqrt{\Lambda_{k}^{2}-(2\Delta_{k})^{2}}$ that
correspond to the excitation spectrum of the $x$-transverse spin triplet
states and $\omega_{k}'=\pm
\sqrt{\Lambda_{k}'^{2}-(2\Delta_{k}')^{2}}$ that correspond to the $y$-transverse one.

\section{Paramagnetic State}
From these modes above and
their bosonic character an expression for the paramagnetic internal
energy at finite temperatures can be easily obtained\cite{D2,D1,D3},

\begin{equation}\label{intpara}
U=\varepsilon_{0}+
\sum_{\mathbf{k}}\left(\omega_{0}n(\omega_0)+ \omega_{\mathbf{k}}n(\omega_{\mathbf{k}})+\omega_{\mathbf{k}}'n(\omega_{\mathbf{k}}')\right)\nonumber\\
\end{equation}
where
$\varepsilon_{0}=N\left(-\frac{3}{4}J\overline{s}^{2}+\mu\overline{s}^{2}-\mu
\right)
+\sum_{\mathbf{k}}(\omega_{\mathbf{k}}+\omega_{\mathbf{k}}'-\Lambda_{\mathbf{k}}-\Lambda_{\mathbf{k}}')/2
$  is the paramagnetic ground state energy, $n(\omega )=\frac{1}{2}\left( \coth\frac{\beta\omega}{2}-1 \right)$ the Bose factor, $\beta=1/k_{B}T$, $k_{B}$ the Boltzman's constant and $T$ the temperature. After some straightforward algebra\cite{D2,D3} using Eq. (\ref{intpara}), the paramagnetic free energy renders
\begin{equation}\label{para}
F=\varepsilon_{0}-\frac{1}{\beta}\sum_{\mathbf{k}}\ln[1+n(\omega_{\mathbf{k}})]-\frac{1}{\beta}\sum_{\mathbf{k}}\ln[1+n(\omega_{\mathbf{k}}')]
-\frac{N}{\beta}\ln[1+n(\omega_{0})].
\end{equation}
For obtaining $\overline{s}^{2}$ and $\mu$ we minimize the free energy by the saddle-point equations
\begin{eqnarray}\label{self}
2(2-\overline{s}^{2})&=&
\frac{1}{2N}\sum_{\mathbf{k}}\left(\frac{\Lambda_{\mathbf{k}}}{\omega_{\mathbf{k}}}\coth\frac{\beta\omega_{\mathbf{k}}}{2}
+\frac{\Lambda_{\mathbf{k}}'}{\omega_{\mathbf{k}}'}\coth\frac{\beta\omega_{\mathbf{k}}'}{2}\right)
+f(\omega_{0}),\nonumber\\
\frac{2J}{t}\left(\frac{3}{4}-\frac{\mu}{J}\right)&=&
\frac{1}{2N}\sum_{\mathbf{k}}\left(\frac{\omega_{0}}{\omega_{\mathbf{k}}}\lambda(\mathbf{k})\coth\frac{\beta\omega_{\mathbf{k}}}{2}
+\frac{\omega_{0}}{\omega_{\mathbf{k}}'}\lambda(\mathbf{k})(1-\delta)\coth\frac{\beta\omega_{\mathbf{k}}'}{2}\right),\nonumber\\
\end{eqnarray}
where $f(\omega_{0})=\frac{N}{2}\left(\coth\frac{\beta\omega_{0}}{2}-1\right)$.
\subsection{Numerical results at $T=0$}
We first study the case $T=0$ e.i., without thermal excitations. At zero temperature the self-consistent equations given by Eqs.
(\ref{self}) can be simplified as,

\begin{eqnarray}\label{T0self}
4(2-\overline{s}^{2})&=&I_{1}(y)+I_{2}(y)+I_{3}(y)+I_{4}(y)\nonumber\\
\frac{4Jy}{t}\left(\frac{3}{4}-\frac{\mu}{J}\right)&=&I_{2}(y)-I_{1}(y)+I_{4}(y)-I_{3}(y),
\end{eqnarray}
with
\begin{eqnarray}
I_{1}(y)&=&\frac{1}{\pi^{d}}\int_{0}^{\pi}\frac{d^{d}k}{\sqrt{1+y\lambda(\mathbf{k})}},\hspace{2mm}
I_{3}(y)=\frac{1}{\pi^{d}}\int_{0}^{\pi}\frac{d^{d}k}{\sqrt{1+y(1-\delta)\lambda(\mathbf{k})}}\nonumber\\
I_{2}(y)&=&\frac{1}{\pi^{d}}\int_{0}^{\pi}d^{d}k\sqrt{1+y\lambda(\mathbf{k})},\hspace{2mm}
I_{4}(y)=\frac{1}{\pi^{d}}\int_{0}^{\pi}d^{d}k\sqrt{1+y(1-\delta)\lambda(\mathbf{k})},
\end{eqnarray}
where we have introduced a dimensionless parameter $y=t\overline{s}^{2}/\omega_{0}$. An equation about $y$ can then be
obtained:
\begin{equation}\label{eqy}
 y=\frac{2t}{J}\left(1-[I_{1}(y)+I_{3}(y)]/4\right).
\end{equation}
We will now obtain the numerical solutions to the zero temperature self-consistent
equations (\ref{T0self}) using Eq. (\ref{eqy}). In this case (paramagnetic phase),
we have the $z$-polarized branch of excitations has a dispersionless
value $\omega_{z}(k)=\omega_{0}$ and the other two branches show a
dispersion which has a minimum at the AF reciprocal vector $Q=(\pi,\pi,\pi)$ in three dimensions ($3d$). The minimum value of  the excitations defines
\begin{eqnarray}\label{spingap}
\Delta^{x}=\omega_{0}\sqrt{1-yd},\hspace{10mm} \Delta^{y}=\omega_{0}\sqrt{1-yd(1-\delta)}.
\end{eqnarray}
The spin gap energy $\Delta^{x}$ and  $\Delta^{y}$ define the energy scale for
the Kondo singlet phase, for $0\leq\delta\leq1$ and $\delta<0$ respectively. For $\delta=0$, $\Delta^{x}$ and $\Delta^{y}$ are identical and we obtain the original spin gap in the KNM\cite{Zhang}. Although we are interested in the case $0<\delta\leq 1$, we consider $\delta<0$ due to theoretical reasons. This case only will be consider at $T=0$ and will not be sketched in this report.
The analysis of the spin gap is important because the vanishing of gap and the appearance of soft modes define the transition from the disordered Kondo spin liquid to the AF phase at the QCP $(J/t=(J/t)_c,T=0)$.
At this point, it is suitable to clarify that in figures (1), (2) and (3), we sketched the spin gap energy like $\Delta/J$ versus $t/J$ by following the $\delta=0$ case\cite{Zhang}, despite we consider throughout this paper the control parameter as $J/t$. That will not yield any physical difference since it only will change the onset of the curves from the left to the right.

\begin{figure}[th]
\label{1d} \centering
\includegraphics[angle=0,scale=0.9]{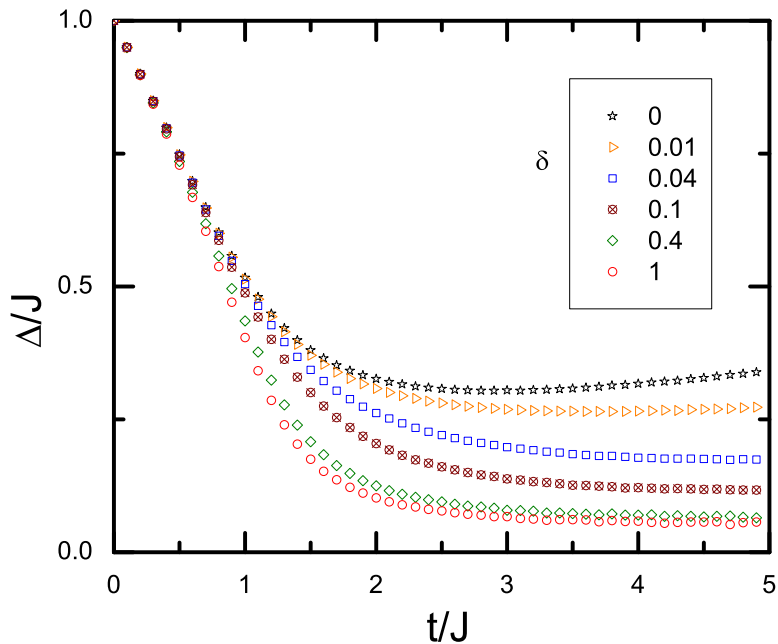}
\caption{(Color online) The spin gap $\Delta/J$ vs the control parameter $t/J$ is sketched for different values of $\delta$ in one dimension and $T=0$. It shows that spin gap is always nonzero for $0\leq\delta\leq1$.}
\end{figure}

\begin{figure}[th]
\label{2d} \centering
\includegraphics[angle=0,scale=0.9]{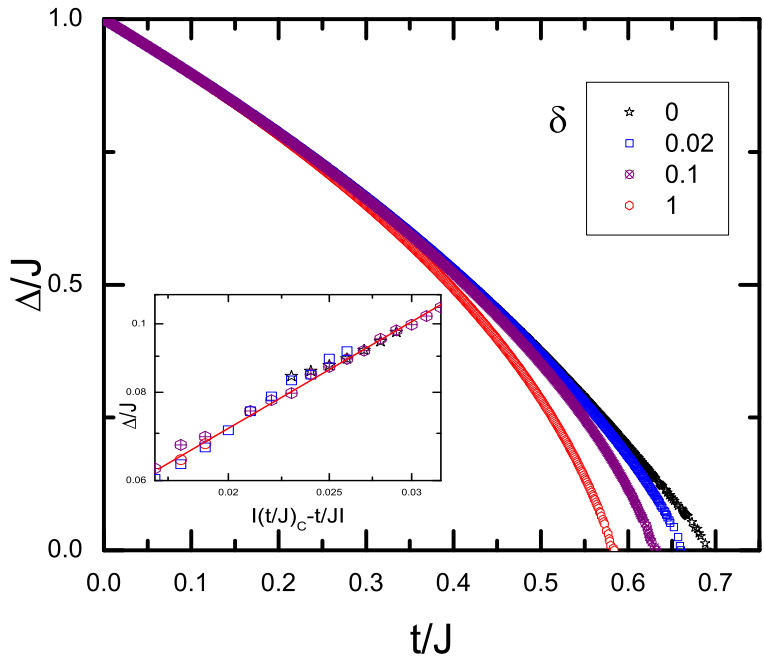}
\caption{(Color online)  Sketch at zero temperature of the spin gap vs the strength $t/J$ in two dimensions. The inset shows the log-log plot of the spin gap versus $|(t/J)_{c}-t/J|$ for $0\leq\delta\leq1$. It shows that $\Delta/J$ vanishes close to $(t/J)_c$ with a exponent $\nu z\approx 1$.}
\end{figure}

\begin{figure}[th]
\label{3d} \centering
\includegraphics[angle=0,scale=0.9]{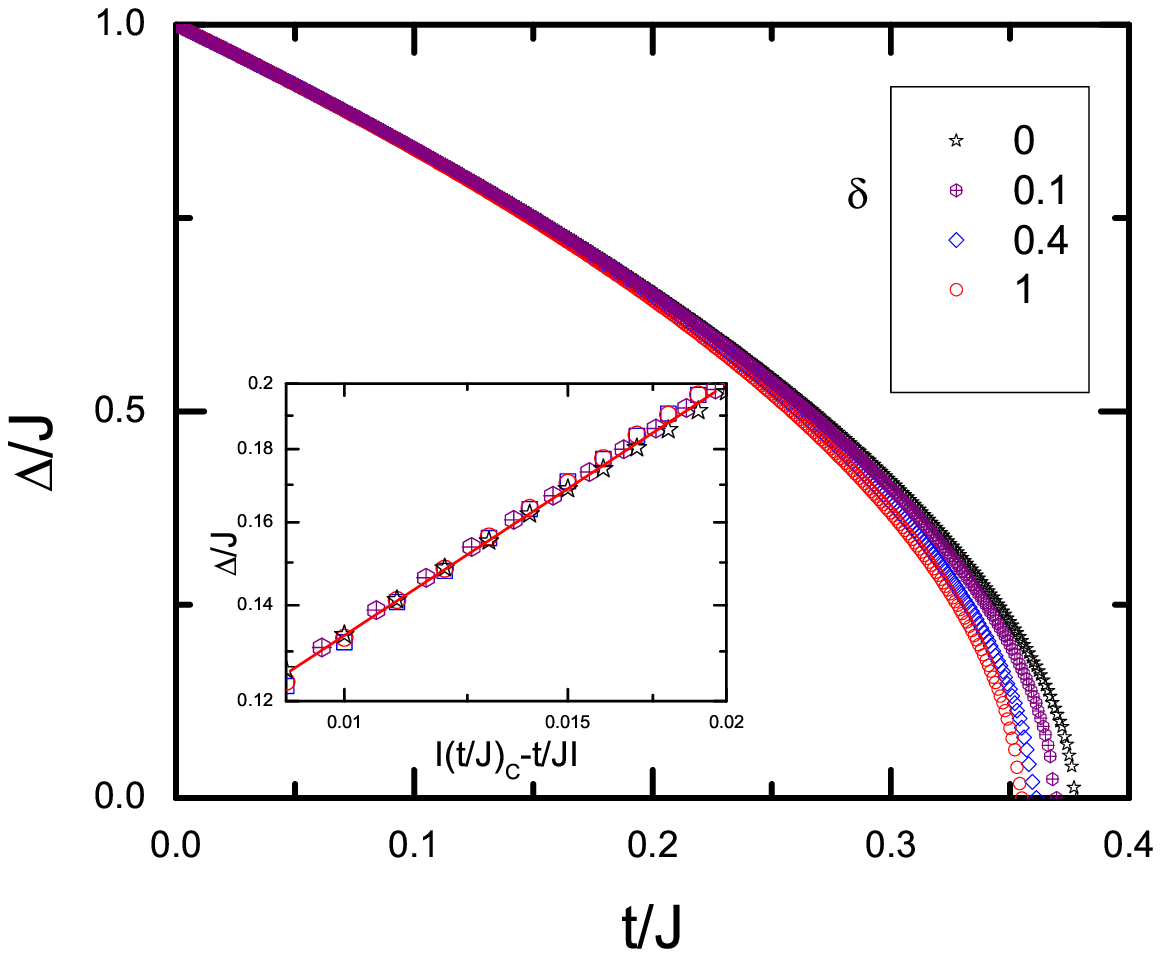}
\caption{(Color online) Anisotropy dependence on the spin gap vs the strength $t/J$ in three dimensions at $T=0$ and $0\leq\delta\leq1$. The scaling of gap close to the QCP is shown in the inset of the figure.  It shows the log-log plot of $\Delta/J$ vs $|(t/J)_c-t/J|$ for $\delta=0,0.1,0.4,1$, and scales close to the QCP like $\Delta/J\sim |(t/J)_c-t/J|^{\nu z}$ with spin gap exponent $\nu z\approx0.5$.}
\end{figure}

In the one dimensional $(1d)$ case, the energy gap falls linearly for small
values of $t/J$ and deviates considerably from the linear behavior as $t/J$ gets larger, as it is relates in Fig (1). Thereby, it is always nonzero for any $\delta$, supporting its disordered phase, own of $1d$ Kondo lattices\cite{Zhang,Tsune}.

The anisotropy dependence on the spin gap in two dimensions ($2d$) is sketched in Fig. (2). For $0<\delta\leq1$ the effect of anisotropy is still weak and it changes the QCP slightly, until $\delta=0$, where both soft modes,  $\Delta^{x}$ and $\Delta^{y}$, contribute and the QCP undergoes a slight jump. Then, the qualitative behavior is the \textit{same} for this range and the gap exponent is approximately $\nu z\simeq1$. It is plotted in the inset of Fig. (2). On the other hand, for $\delta<0$, the QCP is reduced and the Kondo spin liquid phase is limited to a narrower region. It is not shown in Fig. (2).

In three dimensions, the effect of anisotropy on the spin gap is similar as in the $2d$ case. The spin gap follows a exponent $\nu z\simeq0.5$ for $0\leq\delta\leq1$ and changes its universality for $\delta<0$, where the spin gap vanishes close to the QCP more faster. As in the $2d$ case, there is a jump for $\delta=0$ which is the result of all soft modes coincide.

We conclude that, for all anisotropy between $0\leq\delta\leq1$, there exists a critical value $(t/J)_c$, where the spin gap vanishes as $\Delta/J\propto|(t/J)_{c}-t/J|^{\nu z}$, and a QPT to the ordered magnetic phase occurs in $2d$ and $3d$ whereas no transition happens in $1d$. This is similar to the results in Ref. \cite{Zhang} for $\delta=0$ and it gives us a kind of universality similar as in the isotropic Kondo lattices\cite{D2,Zhang,jure}. From relation between the spin gap and the distance to the QCP,  sketched in the onset of Fig. (3), it is shown that when $t/J$ increases from its strong coupling limit, the triplet spin gap at the wave vector $Q=(\pi,\pi,\pi)$ decreases and vanishes at $t/J=(t/J)_c$. Since $\Delta/J\propto|(t/J)_{c}-t/J|^{0.5}$, close to the QPT, we can immediately identify the spin gap exponent $\nu z\approx0.5$ at the QCP of the Kondo lattice, confirming our early theoretical results\cite{D2}. Finally, for $\delta<0$ exists also a QPT in $d=2,3$ but no phase transition appears in $1d$.

\subsection{Analytical results at the quantum critical trajectory}\label{analy}
Since quantum phase transitions are generally associated with soft modes at the QCP, where
the gap for excitation vanishes, then physical quantities have power law temperature
dependencies determined by the quantum critical exponents\cite{livroM}; one of them is the specific heat $C_V$, that we will calculate here. This strategy has been intensively explored in the study of heavy fermion materials, in the so-called \textit{quantum critical trajectory}  $J/t=(J/t)_c$, $T\rightarrow0$, fixing the pressure (in our case the control parameter $J/t$) at its critical value for the
disappearance of magnetic order\cite{Stewart}.
Then, we calculate analytically, the anisotropy dependence on the specific heat at $J/t=(J/t)_{c}$, $T\rightarrow0$ for both cases, $\delta\ll 1$ and $\delta\approx 1$. All the
calculations will be done considering two essential
approximations: $(i)$ The system is at the
quantum critical point  $J/t=(J/t)_{c}$, and temperatures $T\rightarrow0$. $(ii)$  The temperatures region where will be found the specific heat will be lower than the Kondo temperature ($T_K$).
We will begin writing $k=Q +q$ and expanding for small $q$: $\lambda(q)=-d+q^{2}/2 + O(q^4)$, this  yields the spectrum of transverse spin triplet excitations as,
\begin{eqnarray}\label{spectrumq}
\omega_{q}\approx\omega_{0}\sqrt{1+y\lambda(q)}&=&\sqrt{\Delta^{2}+D q^{2}},\nonumber\\
\omega_{q}'\approx\omega_{0}\sqrt{1+y\lambda(q)(1-\delta)}&=&\sqrt{\Delta^{2}+D(1-\delta) q^{2}+\omega_{0}^{2}\delta},
\end{eqnarray}
where $\Delta=\Delta^{x}$ is the spin gap energy given by Eq. (\ref{spingap}) since $0\leq\delta\leq1$, $D=\omega_{0}^{2}/2d$ the spin-wave stiffness at $T=0$, and $\omega_{0}$ is the $z$-polarized dispersionless branch of
excitations. Considering $\Delta=0$, at the QCP\cite{livroM} in the spectrum excitations Eq. (\ref{spectrumq}), and using
$C_{V}=-T\partial^2 F/\partial T^2$ in Eq. (\ref{para}) we get

\begin{equation}\label{heat}
C_V=\frac{S_d}{4k_{B}T^{2}\pi^{d}}\int_{0}^{\pi} dq q^{d-1}(\omega_{q}^{2}+\omega_{q}'^{2})(\sinh^{-2} {\frac{\beta\omega_{q}}{2}}+\sinh^{-2} {\frac{\beta\omega_{q}'}{2}}),
\end{equation}
where $S_d$ is the solid angle. Equation (\ref{heat}) yields the
expression for the anisotropic dependence on specific heat at the  \textit{quantum critical trajectory}, as an contribution of bosons $t_x$ and $t_y$.

\emph{Case $0\leq\delta\ll1$}---Having shown the relationship between the specific heat $C_V$ and $\delta$ we now discuss the case $\delta\ll1$.
Making change of variables in Eq. (\ref{heat}) we obtain,
\begin{equation}\label{intab}
C_{V}(\delta\ll1)=\frac{S_{d}k_{B}Z^{d/2}}{\pi^{d}}\left(\frac{k_{B}T}{\omega_0}\right)^{d}[\Upsilon_{1}(d)+\frac{\delta}{4}(\Upsilon_{2}(d)-2\Upsilon_{1}(d))],
\end{equation}
where $\Upsilon_{1}(d)=\int_{0}^{\infty}dx x^{d+1}\sinh^{-2}(x/2)$, $\Upsilon_{2}(d)=\int_{0}^{\infty}dx x^{d+2}\coth(x/2) \sinh^{-2}(x/2)$ and
$x=\beta\omega_{0}q/\sqrt{Z}$. In two dimensions we found $\Upsilon_{1}(2)=24\zeta(3)$ and $\Upsilon_{2}(2)=96\zeta(3)$, where $\zeta$ is the Riemann zeta-function. In three dimensions  $\Upsilon_{1}(3)=16\pi^{4}/15$ and $\Upsilon_{2}(3)=16\pi^{4}/3$. For $\delta=0$, the spectrum excitations given by Eq. (\ref{spectrumq}) coincide and we recover the exact value as
obtained in an previous work for the isotropic KNM\cite{D2}.

\emph{Case $\delta\approx 1$}---Here, it is sufficient to consider, $\xi=1-\delta\ll 1$,
where $\xi$ is a dimensionless parameter that controls the Ising-like anisotropy in this case.
Thereby, working in analogy with the preceding case, we obtain
\begin{equation}\label{intab}
C_V(\delta\approx1)=\frac{S_{d}k_{B}Z^{d/2}}{4\pi^{d}}\left(\frac{k_{B}T}{\omega_0}\right)^{d}\Upsilon_{1}(d)(2-\delta),
\end{equation}
where we have already replaced the $\xi$ expression.
The results above show that the specific heat of the AKNM for $\delta\ll1$ and $\delta\approx1$ is only re-normalized by the
anisotropy, concluding that  $C_V\propto T^{d}$ at the \textit{quantum critical trajectory} for $\delta\ll1$ and $\delta\approx1$. Notice
that this is consistent with the general scaling result $C_V
\propto T^{d/z}$ with the dynamic exponent taking the value
 $z=1$\cite{livroM}. Since $z=1$, in three dimensions
$d_{eff}=d+z=d_c=4$ where $d_c$ is the upper critical dimension
for the magnetic transition \cite{livroM}. Consequently, the
present approach yields the correct description of the quantum
critical point of the Kondo lattices for $d\ge 3$.

\section{Antiferromagnetic Phase}

The mean-field approach can be extended to the AF phase assuming the
condensation in the $x$ component of the spin triplet like:
$t_{\mathbf{k},
x}=\sqrt{N}\bar{t}\mathbf{\delta_{k,Q}}+\mathbf{\eta}_{\mathbf{k},x}$, where $\bar{t}$ is its mean value in the ground state and $\mathbf{\eta}_{\mathbf{k},x}$ represents the fluctuations. Making the same steps as before, the internal energy  renders
\begin{equation}\label{free}
U'=\varepsilon_{0}'+
\sum_{\mathbf{k}}\left(\omega_{0}n(\omega_0)+ \omega_{\mathbf{k}}n(\omega_{\mathbf{k}})+\omega_{\mathbf{k}}'n(\omega_{\mathbf{k}}')\right),\nonumber\\
\end{equation}
where $\varepsilon_{0}'=N\left[-\frac{3}{4}J\overline{s}^{2}+\mu\overline{s}^{2}-\mu+\left(\frac{J}{4}+\mu-
\frac{1}{2}tZ\overline{s}^{2}\right)\overline{t}^{2} \right]
+\sum_{\mathbf{k}}(\omega_{\mathbf{k}}+\omega_{\mathbf{k}}'-\Lambda_{\mathbf{k}}-\Lambda_{\mathbf{k}}')/2$ is the AF ground state. The free energy is now
\begin{equation}\label{Gibbs}
F'=\varepsilon_{0}'-\frac{1}{\beta}\sum_{\mathbf{k}}\ln[1+n(\omega_{\mathbf{k}})]-\frac{1}{\beta}\sum_{\mathbf{k}}\ln[1+n(\omega_{\mathbf{k}}')]
-\frac{N}{\beta}\ln[1+n(\omega_{0})].
\end{equation}
Minimizing the free energy Eq. (\ref{Gibbs}), using
$(\partial F'/\partial\mu,\partial F'/\partial\overline{s},\partial F'/\partial \bar{t})=(0,0,0)$,
we can easily get the following saddle-point equations,
\begin{eqnarray}\label{st}
\overline{s}^{2}&=&1+\frac{J}{Z
t}-\frac{f(\omega_0)}{2}\nonumber\\
&-&\frac{1}{4N}\sum_{\mathbf{k}}\left(\sqrt{1+\frac{2\lambda(\mathbf{k})}{Z}}(1+2n(\omega_{\mathbf{k}}))
+\sqrt{1+\frac{2\lambda(\mathbf{k})(1-\delta)}{Z}}(1+2n(\omega_{\mathbf{k}}'))\right), \nonumber
\\
\overline{t}^{2}&=&1-\frac{J}{Z
t}-\frac{f(\omega_0)}{2}-\frac{1}{4N}\sum_{\mathbf{k}}\left(\frac{(1+2n(\omega_{\mathbf{k}}))}{\sqrt{1+\frac{2\lambda(\mathbf{k})}{Z}}}
+\frac{(1+2n(\omega_{\mathbf{k}}'))}{\sqrt{1+\frac{2\lambda(\mathbf{k})(1-\delta)}{Z}}} \right)
,\nonumber\\
\mu&=&\frac{1}{2}Zt\overline{s}^{2}-J/4,
\end{eqnarray}
with the excitations spectrum of the $x$-transverse and $y$-transverse spin triplet
states given now by, $\omega_{{\bf k}}=\frac{1}{2}Zt\overline{s}^{2}\sqrt{1+2\lambda(\mathbf{k})/Z }$ and $\omega_{{\bf
k}}'=\frac{1}{2}Zt\overline{s}^{2}\sqrt{1+2\lambda(\mathbf{k})(1-\delta)/Z }$, respectively. Generally the equations for $\overline{s}$ and $\overline{t}$ in Eq.
(\ref{st}) should be solved and for $\delta=0$ the results of Ref. (\cite{D2}) are
recovered. Here, in the magnetic ordered state, the condensation of triplets (singlets) follows
from the RKKY interaction (Kondo effect). At finite temperatures the condensation of singlets
occurs at a temperature scale which, to a first approximation,
tracks the exchange $J$ while the energy scale below which the triplet
excitations condense is given by the critical Neel temperature ($T_N$) which
is calculated in the next section. Thus, the fact that at the
mean-field level, both $\overline{s}$ and $\overline{t}$ do not
vanish may be interpreted as the coexistence of Kondo screening and
antiferromagnetism in the ordered phase\cite{D2,Zhang,jure} for all
values of the ratio $J/t<(J/t)_c$.

\section{Critical line in the AKNM}
Following the discussion above, the critical line giving the finite temperature instability of the
AF phase for $J/t <(J/t)_c$ is obtained making $\overline{t}=0$. Hence, from Eq. (\ref{st}) we can obtain
the boundary of the AF state as,
\begin{eqnarray}\label{g}
\frac{|g|}{Z}&=&\frac{1}{2N}\sum_{\mathbf{k}}\left(\frac{n(\omega_{\mathbf{k}})}{\sqrt{1+\frac{2\lambda(\mathbf{k})}{Z}}}
+
\frac{n(\omega_{\mathbf{k}}')}{\sqrt{1+\frac{2\lambda(\mathbf{k})(1-\delta)}{Z}}}\right)
+\frac{f(\omega_0)}{2},
\end{eqnarray}
where $g=|(J/t)_{c}-(J/t)|$ measures the distance to the QCP. The
latter is given by,
$(J/t)_{c}=Z[1-\frac{1}{4N}\sum_{\mathbf{k}}(\frac{1}{\sqrt{1+2\lambda(\mathbf{k})/Z}}+\frac{1}{\sqrt{1+2\lambda(\mathbf{k})(1-\delta)/Z}})]$, which
separates an antiferromagnetic long range ordered phase from a gapped spin liquid phase. Performing the same analysis as in sub-section (\ref{analy}), expanding the spectrum excitations close to $\mathbf{Q}=(\pi,\pi,\pi)$,  Eq. (\ref{g}) becomes
\begin{equation}\label{gz}
\frac{|g|}{Z}=\frac{S_d\omega_{0}}{4\pi^{d}}\int_{0}^{\pi} dq q^{d-1}\left(\frac{1}{\omega_{q}}\left(\coth {\frac{\beta\omega_{q}}{2}}-1\right)+\frac{1}{\omega_{q}'}\left(\coth {\frac{\beta\omega_{q}'}{2}}-1\right)\right),
\end{equation}
where we have considered that for temperatures $k_{B}T\ll\omega_{0}$, $f(\omega_0)$ goes to zero faster
than the first term of Eq. (\ref{g}). This equation above allow us to obtain the critical line in the AKNM as a function of the anisotropy parameter $\delta$.

\subsection{Case $0\leq\delta\ll 1$}

We now demonstrate analytically the appearance of a finite Neel
line temperature when a small degree of anisotropy $\delta$ in $y$-component spin
is turned on. Then, solving Eq. (\ref{gz}) for $0\leq\delta\ll1$, we get
\begin{eqnarray}\label{int}
\frac{|g|}{Z}_{\delta\ll1}=\frac{S_{d}Z^{d/2}}{2\pi^{d}}\left(\frac{k_{B}T}{\omega_{0}}\right)^{d-1}\left[\Phi_{1}(d)+\frac{\delta}{8}(\Phi_{2}(d)+2\Phi_{1}(d))\right],
\end{eqnarray}
where $\Phi_{1}(d)=\int_{0}^{\infty}dxx^{d-2}\left(\coth \frac{x}{2}-1\right)$ and $\Phi_{2}(d)=\int_{0}^{\infty}dx x^{d+1}\sinh^{-2}(x/2)$. We notice that the integrals $\Phi_{1}(d)$ and $\Phi_{2}(d)$ diverge for $d<3$
showing that there is no critical line in two dimensions at finite
temperatures\cite{D2,D1} for any anisotropy $\delta\ll1$, in agreement with the Mermin-Wagner
theorem\cite{Mermin}. Nevertheless, for $d\geq3$, the integrals are finite and the
equation for the critical line shows, $(T_{N})_{\delta\ll 1}\propto|g|^{\phi}$, with $\phi = 1/(d-1)$.
If we write the equation for the critical line, $f(g,T)=0$, in the form,
$(J/t)_{c}(T)-(J/t)_{c}(0)+v_0T^{1/\psi}=0$, with $v_0$ related to the
spin-wave interaction, we identify the {\em shift exponent},
$\psi=z/(d+z-2)$\cite{Millis}, that comparing with $\phi$ gives us the dynamic exponent $z=1$, a Gaussian result, since the critical line
only exists for $d>2$. The temperature dependence of the function $f$
arising from the spin-wave interactions can modify the temperature
dependence of the physical properties, as the specific heat,  at $J/t=(J/t)_{c}$. However, in the
limit $T \rightarrow 0$ we can easily see that the purely Gaussian
results for the specific heat calculated in section (\ref{analy}) is dominant, in concordance with the mean-field treatment used here.
For $\delta=0$, we obtain the well established result for the critical line in the KNM\cite{D2}, this due to the fact that the spectrum energy of the two excitations coincide. For $\delta\approx1$, following the same steps as before, it is
straightforward to show that the dominant is also $T_{N}(\delta\approx1) \propto |g|^{\phi}$ for $d\geq3 $, and no critical line exists for $d=2$.

In summary, we have obtained analytically the expression for the Neel line, below which the triplet excitations condense, close to
the QCP for $0\leq\delta\ll1$ and $\delta\approx1$. We have shown that this line does not exist for $d=2$ for any value of the anisotropy, as we expected, whereas for $d\geq3$, the power dependence on $|g|$ of the critical line in the presence of the anisotropy is the \textit{same} of the KNM original. Therefore, the criticality close the QCP is governed by the \textit{same} critical exponents of the isotropic $\delta=0$ case that we have calculated before\cite{D2}.

\section{Conclusions}
In conclusion, we have examined the phase diagram of the Kondo necklace model
in the presence of an Ising-like anisotropy at zero and low temperatures by means of analytical and numerical techniques.
At zero temperature we have derived and solved the self-consistent equations
on the Kondo spin liquid phase for any value of $\delta$.
This allows us to calculate the anisotropy dependence on the spin gap for $d=1,2,3$.
In the $1d$ case, there is no indication at all suggesting a critical value for $t/J$ where
the gap would vanish for any value of anisotropy $\delta$. For $d=2,3$ we found that
the anisotropy in the range $0<\delta\leq1$ dislocates lightly the QCP until $\delta=0$,
where the spectrum excitations coincide. In this range $0\leq\delta\leq1$ the spin gap exponent
is approximately the \textit{same}, while  for $\delta<0$ a like-jump occurs and it belongs to
other universality class. In particular, in three dimensions, the triplet spin gap for anisotropy
$0\leq\delta\leq1$, close to the wave vector $Q=(\pi,\pi,\pi)$, decreases and vanishes at
$t/J=(t/J)_c$ with spin gap exponent $\nu z\approx0.5$, consistent with the dynamic exponent $z=1$
and correlation length $\nu=1/2$, a result in agreement with the mean-field or Gaussian
character of the approximations we have used to deal with the bond-operator Hamiltonian.
On the other hand, at low but finite temperatures, we find that in general the dependence
on $|g|$ of the critical line for the AKNM, in the presence of the anisotropy, is the \textit{same}
of the KNM original. This implies that the critical exponents controlling the transition close to the QCP,
for nonzero $\delta$, are the same as those of the isotropic case. We have also obtained the
thermodynamic behavior of the specific heat along the \textit{quantum critical trajectory}
$J/t=(J/t)_c$, $T\rightarrow0$. It has a power law temperature dependence as $C_V\propto T^{d}$,
a result consistent with the scaling theory with the dynamic exponent $z=1$.
Therefore, the most essential features of the Kondo lattices, i.e., the competition between a
long-range-ordered state and a disordered state is clearly retained in the model for
$0\leq\delta\leq1$. The qualitative features regarding the stability of the AF phase are
well displayed in the model and it allows a simple physical
interpretation of the phase diagram in anisotropic Kondo lattices.
It will be left to a further work to compare our theoretical
results obtained for the AKNM with experimental data in order to clarify to what
extent the estimates of $\delta$ from measured quantities depend on
the theoretical tools used.

\subsection{Acknowledgments}
D. Reyes to thank professor Andre M. C. de Souza for useful computational help.
The authors would like to thank also the Brazilian Agency, CNPq
for financial support.

\end{document}